# Magnetism in Gallium doped CeFe$_2$: The martensitic scenario

Arabinda Haldar[1], K. G. Suresh[1*] and A. K. Nigam[2]
[1]Magnetic Materials Laboratory, Department of Physics,
Indian Institute of Technology Bombay, Mumbai- 400076, India
[2]DCMPS, Tata Institute of Fundamental Research,
Homi Bhabha Road, Mumbai- 400001, India

*Abstract*

Ce(Fe$_{1-x}$Ga$_x$)$_2$ compounds with $x$=0, 0.01, 0.025 and 0.05 have been investigated to unravel the effect of Ga on the magnetic state of CeFe$_2$. For the first time, we find that the dynamic antiferromagnetic phase present in CeFe$_2$ gets stabilized with Ga substitution. The hysteresis loops show that while the compounds with $x$=0 and 0.01 show normal behavior, the other two show multiple magnetization steps across the antiferromagnetic-ferromagnetic transition region. The virgin curve is found to lie outside the envelope curve in these two compounds, similar to the observations made in Ru and Re substituted CeFe$_2$ compounds. Temperature, sweep rate and time dependences of the magnetization show that the compounds with $x \geq 0.025$ possess glassy behavior at low temperatures. Various results obtained reveal that these two compounds belong to the martensite family.

-----------------------------------------------------------------
*Corresponding author (email: suresh@phy.iitb.ac.in)

## I. Introduction

Rare earth (*R*) -iron phase with the composition $R\text{Fe}_2$ generally crystallize in the fcc Laves-phase structure and are simple ferro- or ferrimagnets[1-3]. Many members of this series have their Curie temperatures ($T_C$) well above the room temperature and many of them show large magnetostriction. The *R*-Fe coupling is known to be ferromagnetic (FM) in the case of light rare earths and antiferromagnetic (AFM) in the case of heavy rare earths. The strong hybridization between Ce 4*f* and Fe 3*d* electrons makes $\text{CeFe}_2$ very special in the $R\text{Fe}_2$ series. Although Ce is a light rare earth, 4*f* electrons hybridize antiferromagnetically with the 3*d* electrons and it is due to the quenching of orbital 4*f* moment by the band formation[1]. Strong anomaly in the lattice parameter is observed in comparison with the smooth decrease through the $\text{RFe}_2$ series, leading to a lattice parameter for $\text{CeFe}_2$ close to that of $\text{HoFe}_2$. The $T_C$=230K is almost lower by a factor of 3 as compared to that of $\text{LuFe}_2$ ($T_C$=610K) with its full 4f shell and $\text{YFe}_2$ ($T_C$=545K) which has no 4f electrons. The saturation magnetic moment is also anomalously low ($M_s$= $2.3\mu_B$ / f.u.) with respect to that of $\text{LuFe}_2$ ($M_s$= $2.9\mu_B$ / f.u.)[2,4]. Another interesting observation in $\text{CeFe}_2$ is the occurrence of antiferromagnetic fluctuations[1].

It has been reported that substitution of small amounts of elements such as Co, Al, Ru, Ir, Os and Re stabilizes the low temperature dynamic AFM phase in $\text{CeFe}_2$ [3,5]. The co-existence of FM and AFM phases across the AFM-FM transition has been shown by Hall probe imaging and this transition bears distinct signatures of first-order phase transition, namely, supercooling, superheating and time relaxation[6]. Sharp change in the magnetization have been reported across the AFM to FM transition at temperatures less than 5 K when $\text{CeFe}_2$ is doped with Ru and Re[7]. Multi-step magnetization behavior is another characteristic of these compounds. This behavior is explained by a disorder-influenced first-order magnetostructural phase transition[7].

Interestingly, these kinds of sharp magnetization steps are observed in mixed-valent manganese oxides with general formula $\text{Pr}_{1-x}\text{Ca}_x\text{Mn}_{1-y}\text{M}_y\text{O}_3$ (with *x*~0.5, *y*~0.05 and where M is the cation used to destabilize the Mn sublattice)[8-10]. Ultrasharp magnetization steps are also found in the intermetallic compound $Gd_5Ge_4$ [11]. Basically, these materials are well known phase-separated systems and the transformation between these two phases has a martensitic character. Detailed studies have been performed on these materials using the field sweep rate and time dependence of magnetization [12-14]. Wu et al [14] have suggested that an induction period exists for these steps to appear, implying that the dynamics of the strain field organization is a critical ingredient behind this phenomenon. Similarities between manganites such as $Pr_{0.6}Ca_{0.4}Mn_{0.96}Ga_{0.04}O_3$ and intermetallic compound $Gd_5Ge_4$ have been established by Hardy et al [12].

Doped $CeFe_2$ alloys are another series of materials which show phase separated magnetic structure while varying field, temperature or pressure. Very few studies have been done to explain these sharp steps in this series of compounds[7]. To examine the deeper interconnection between the phase separation and magnetism in these materials, we made a detailed magnetization study on Ga doped $CeFe_2$ alloys and the results are presented in this paper. To the best of our knowledge no report on sweep rate dependence or time delay measurement in doped $CeFe_2$ alloys has been reported. This prompted us to focus our attention on the influence of these two variables on the magnetization behavior.

## II. Experimental Details

Polycrystalline samples of $Ce(Fe_{1-x}Ga_x)_2$ [ $x$=0, 0.01. 0.025 and 0.05] were prepared by arc melting the stoichiometric proportion of the constituent elements of at least 99.9% purity, in a water cooled copper hearth in purified argon atmosphere. The resulting ingots were turned upside down and remelted several times to ensure homogeneity. The weight loss was monitored at the end of the melting process and the characterization was performed only on samples whose final weight loss was less than 0.5%. The as-cast samples were annealed at 600 ºC for 2 days, 700 ºC for 5 days, 800 ºC for 2 days and 850 ºC for 1 day[3]. The structural analysis of the samples was performed by collecting the room temperature powder x-ray diffractograms (XRD) using Cu-K$\alpha$ radiation. The refinement of the diffarctograms was done by the Rietveld analysis using *Fullprof* suite program. The lattice parameters were calculated from the refinement. The DC magnetization measurements in the temperature range of 1.8- 300 K and in fields up to 90 kOe were performed with the help of Physical Property Measurement System (PPMS, Quantum Design Model 6500) which has a vibrating sample magnetometer (VSM) attachment. Some measurements were done using Oxford Maglab VSM.

## III. Results

Fig. 1 shows the room temperature powder x-ray diffraction pattern of $Ce(Fe_{1-x}Ga_x)_2$ compounds along with the Rietveld refinement. The difference plot between the theoretical and the experimentally observed patterns is shown at the bottom of each plot. The refinement confirms that all these compounds are single phase, crystallizing in the $MgCu_2$ type cubic structure, with the space group of FD3m. The lattice parameters obtained from refinement are 7.3018(3) for x=0, 7.3059(3) for x=0.01, 7.3090(3) for x=0.025 and 7.3097(5) Å for x=0.05.

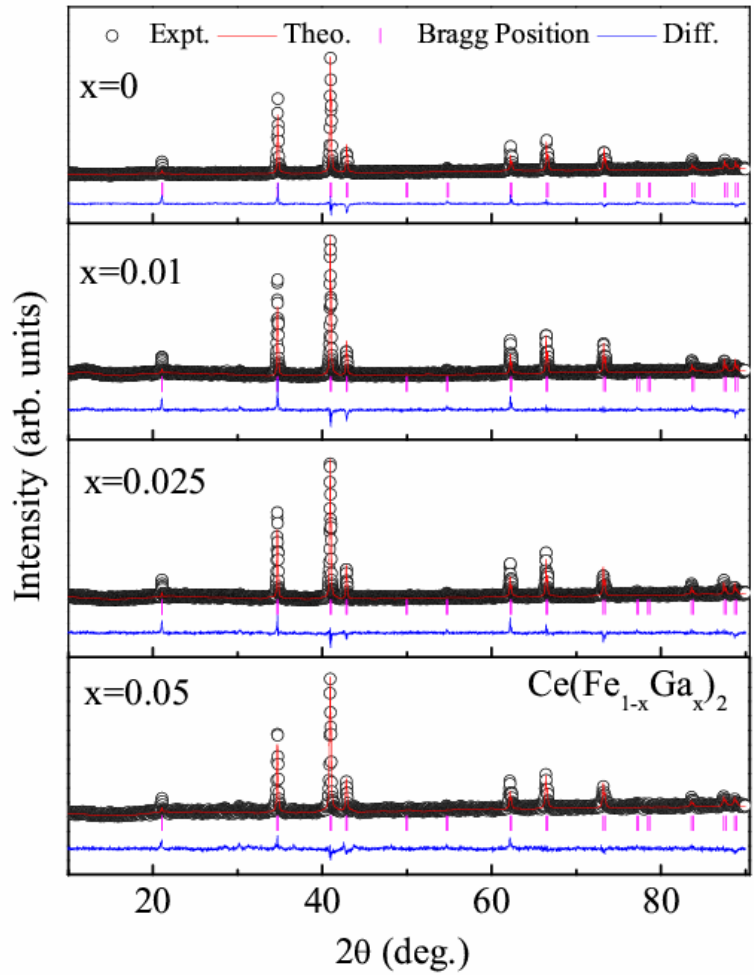

FIG. 1. Powder x-ray diffraction patterns, along with the Rietveld refinement of Ce(Fe$_{1-x}$Ga$_x$)$_2$ compounds. The plots at the bottom show the difference between the theoretical and the experimental data.

The temperature ($T$) variation of magnetization ($M$) of Ce(Fe$_{1-x}$Ga$_x$)$_2$ compounds has been studied in an applied field of 500 Oe both under 'zero-field cooled' (ZFC) and 'field-cooled warming' (FCW) conditions. In both the modes, the data was collected during the warming cycle. In all the compounds, at low fields, both ZFC and FCW data follow almost the same path as the temperature is varied. Fig. 2a shows the $M$ vs. $T$ plots of Ce(Fe$_{1-x}$Ga$_x$)$_2$ samples. The high temperature transition corresponds to the FM-paramagnetic (PM) transition. It can be seen that for x≥0.025, the unstable antiferromagnetic state in CeFe$_2$ gets stabilized, as indicated by the low temperature transition. Furthermore, upon Ga substitution, the $T_C$ shows a decreasing trend even with low concentrations of Ga. Fig. 2b shows the typical field variation of the *M-T* plots for the compound with *x*=0.025. The antiferromagnetic transition gets gradually suppressed by the field and at 40 kOe, it is almost completely suppressed. Analysis of the temperature variation of magnetization has shown that the magnetization at low temperature is dictated predominantly by the spin wave excitations than the Stoner excitation.

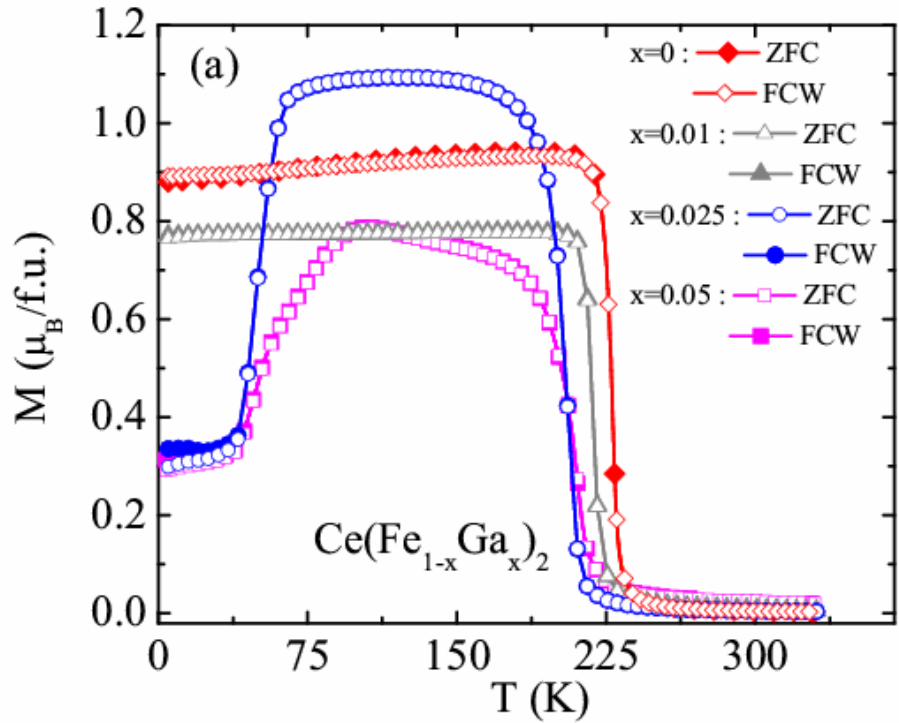

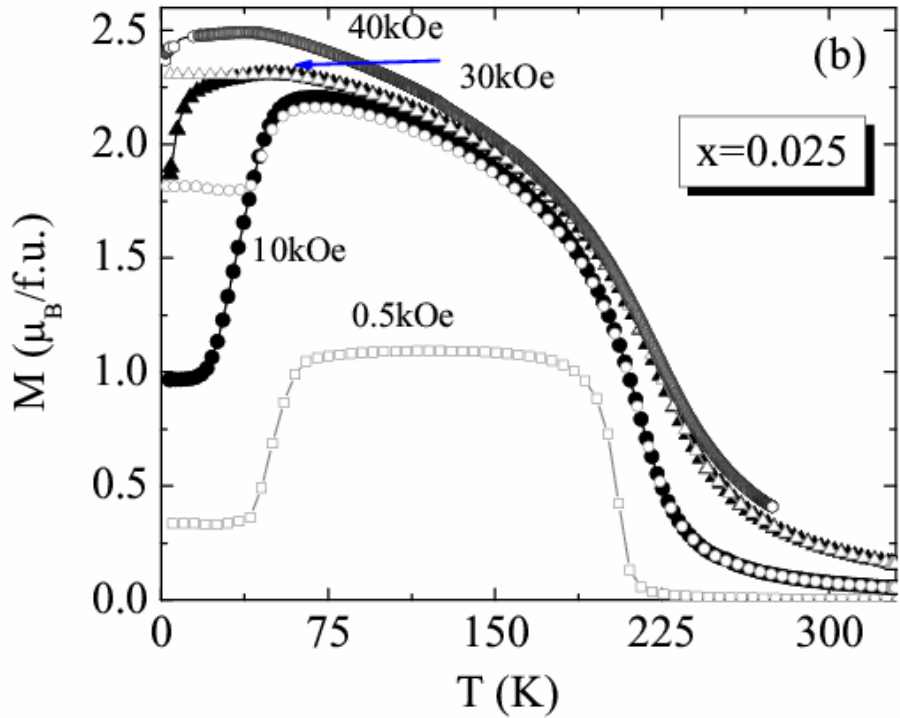

FIG. 2. Temperature dependence of the ZFC magnetization of (a) Ce(Fe$_{1-x}$Ga$_x$)$_2$ compounds at 500 Oe, (b) Ce(Fe$_{0.975}$Ga$_{0.025}$)$_2$ in various fields.

Fig. 3a-c shows the isothermal magnetization curves below 3K for various concentrations of Ga doped compounds. All the measurements have been made with a field sweep rate of 100 Oe per second. The samples were zero-field cooled to the measurement temperature. The set temperature was almost constant throughout the measurement time. While the compounds with $x=0$ and 0.01 show the normal ferromagnetic behavior, those with $x=0.025$ and 0.05 show very interesting behavior. Fig. 3a shows the *M-H* isotherms taken for CeFe$_2$ and Ce(Fe$_{0.99}$Ga$_{0.01}$)$_2$ compounds at 2 K in the increasing and decreasing field cycles. Both these compounds show normal ferromagnetic behavior without any anomaly. However, the compounds with $x=0.025$ and 0.05 show distinct jumps in the magnetization curves. We define some critical field $H_c$ at which *M-H* curve changes the slope abruptly. It is important to note that the value of $H_c$ will depend on the details of measurement procedure and the thermal and magnetic history of the sample.

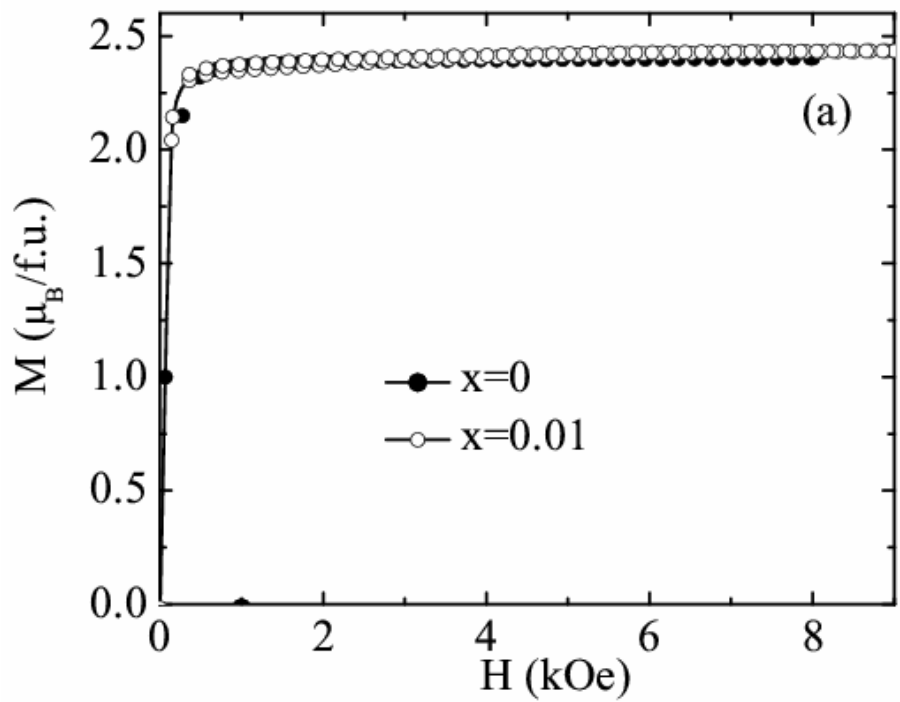

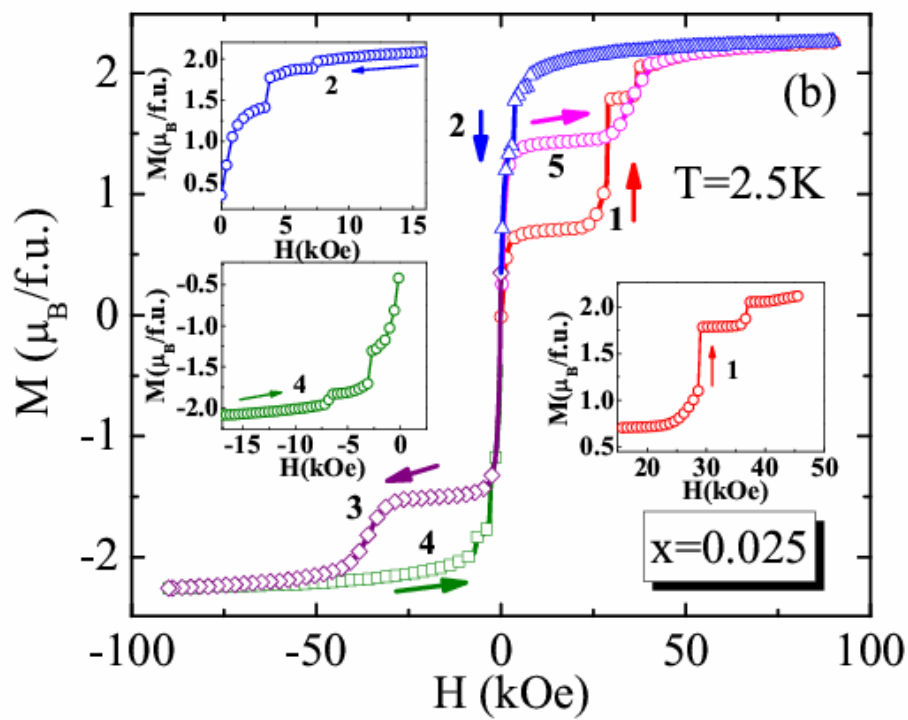

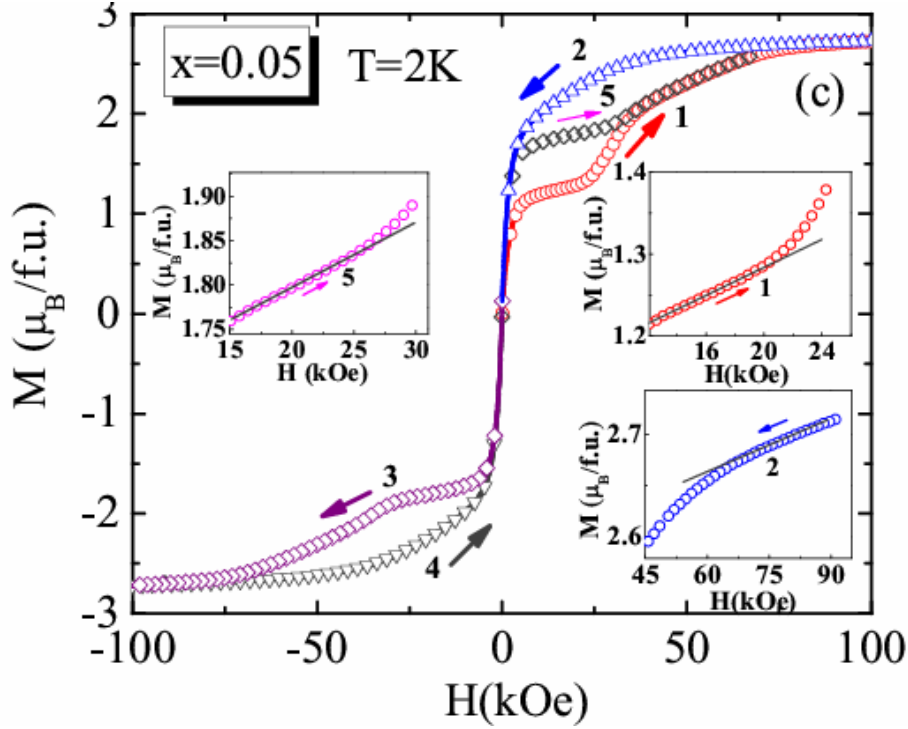

FIG. 3. *M-H* isotherms of (a) $CeFe_2$ and $Ce(Fe_{0.99}Ga_{0.01})_2$ at 2 K, (b) $Ce(Fe_{0.975}Ga_{0.025})_2$ at 2.5 K and (c) $Ce(Fe_{0.95}Ga_{0.05})_2$ at 2 K. The samples have been cooled in zero field to the measurement temperature. The arrows indicate the directions of field change. The insets in (b) and (c) highlight the abrupt changes along various paths.

We have already seen that the fluctuating AFM state becomes stable in the compounds with $x$=0.025 sample and 0.05 below 50K (Fig. 2a). Five quadrant *M-H* isotherms have been taken on these samples at temperatures below 3 K. The AFM phase gets converted to FM phase during the field increment from 0 to 90kOe, giving rise to sharp multiple transitions in this path (path 1, inset of Fig. 3b) in the $x$=0.025 sample, whereas comparatively smooth and single transition (at $H_{c1}$=20kOe) is observed in x=0.05 sample (path 1, inset of Fig. 3c). Two sharp steps are observed, one at $H_{c1}$ ~ 24kOe and another step is found at $H'_{c1}$ ~ 36kOe (path1, inset of Fig. 3b) in $x$=0.025 sample. It has been reported that Ru and Re doped $CeFe_2$ samples show similar sharp magnetization step across the field induced AFM-FM transition when the measurement is performed below 5K[7]. This is due to the intrinsic canted structure of AFM state[15,16].

When the field is reduced, the system again goes to AFM state but in a relatively gradual way. In $x$=0.025 sample some small steps are observed in the path 2 (path 2, inset of Fig.3b) which is absent in x=0.05 sample. The growth of AFM phase with the decrease of $H$ occurs at ~45kOe in the compound with x=0.025 (not shown in the inset) and at ~72kOe for x=0.05 sample (path 2, inset of Fig. 3c). These fields are larger than the fields where the AFM to FM transition occurs, in both the compounds. The difference in the transition process in the increasing and decreasing field cycles in doped CeFe$_2$ samples has earlier been attributed to the asymmetry between supercooling and superheating across a first order phase transition[17]. This kind asymmetry is absent in CMR-manganites and Gd$_5$Ge$_4$, though there are some similarities between these systems [18]. When the field is reversed (path 3), the transition from AFM to FM state occurs at ~ 25kOe for both x=0.025 and 0.05. These values are slightly higher than the values obtained when data was taken along the virgin curve (path 1). When the field is increased to zero, the system again goes to AFM state at almost same transition fields as in path 2 in both the samples. Interestingly, on further increase of the field, the envelope curve lies inside the virgin curve. If we concentrate on paths 1 and 5, it can be seen that they are very much different from each other. In $x$=0.025 sample, there is no sharp step found along the envelope curve whereas two steps are observed along the virgin curve. It can also be seen that along the envelope curve, the AFM to FM transition takes place at ~ 25kOe for x=0.025 sample and ~ 28kOe for x=0.05 sample.

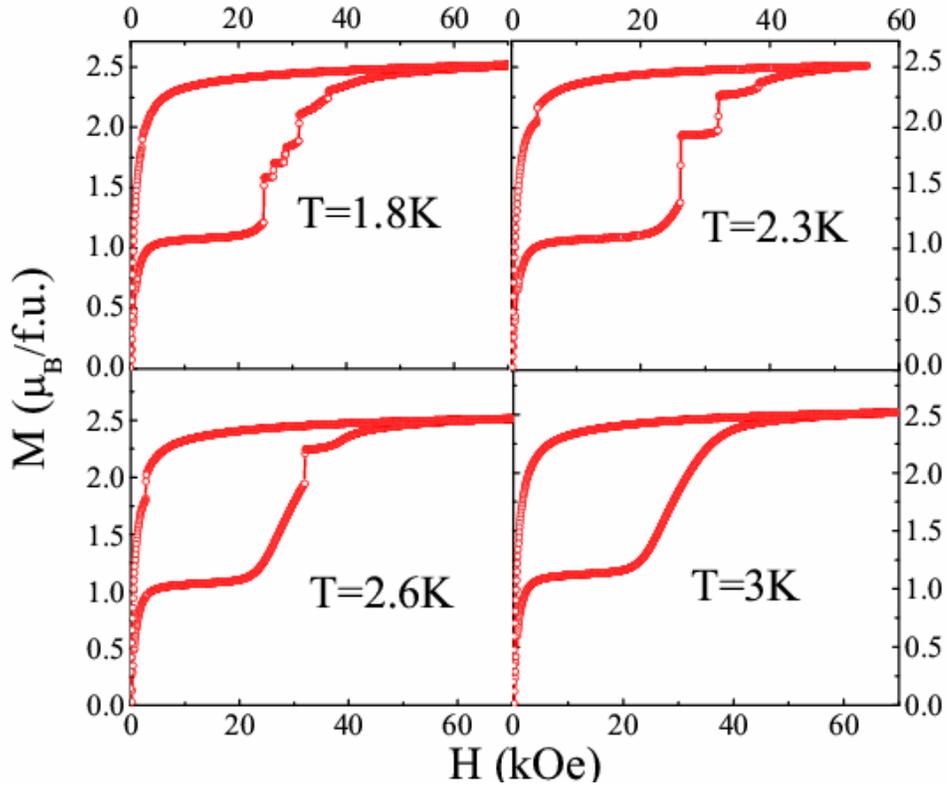

FIG. 4. Two loop magnetization isotherms for Ce(Fe$_{0.975}$Ga$_{0.025}$)$_2$ at various temperatures below 3 K. The sample was cooled in zero field from 240K before each measurement.

Fig. 4 shows the magnetization isotherms taken at different temperature below 3K. All the measurements have been performed after cooling the sample from 240K (>T$_C$). At 3 K, a smooth phase transition is observed. When the temperature is reduced to 2.6 K a step type transition is observed. With further reduction in temperature the *M-H* isotherms consist of a number of ultrasharp steps before it transforms to fully ferromagnetic phase. The M-H curves were also recorded without thermal cycling. Fig. 5 shows the two loop magnetization isotherms for Ce(Fe$_{0.975}$Ga$_{0.025}$)$_2$ without thermal cycling. The data was taken as the field was varied in the sequence 0-90-0-65-0 kOe. It can be seen that there is a large difference between the first and the second runs.

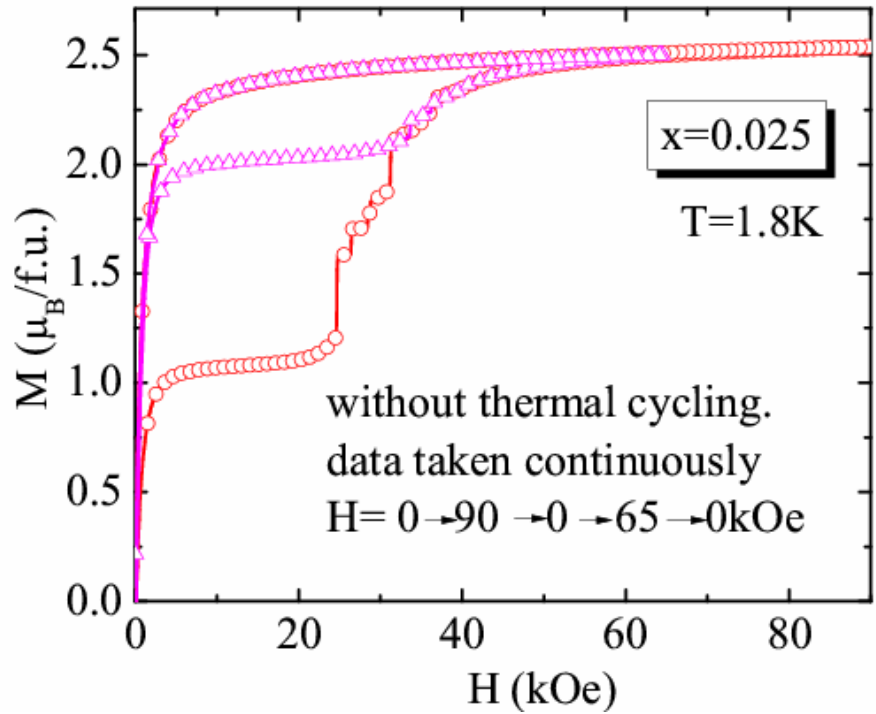

FIG. 5. Two loop magnetization isotherms for Ce(Fe$_{0.975}$Ga$_{0.025}$)$_2$ without thermal cycling. The data was taken in the sequence 0-90-0-65-0 kOe.

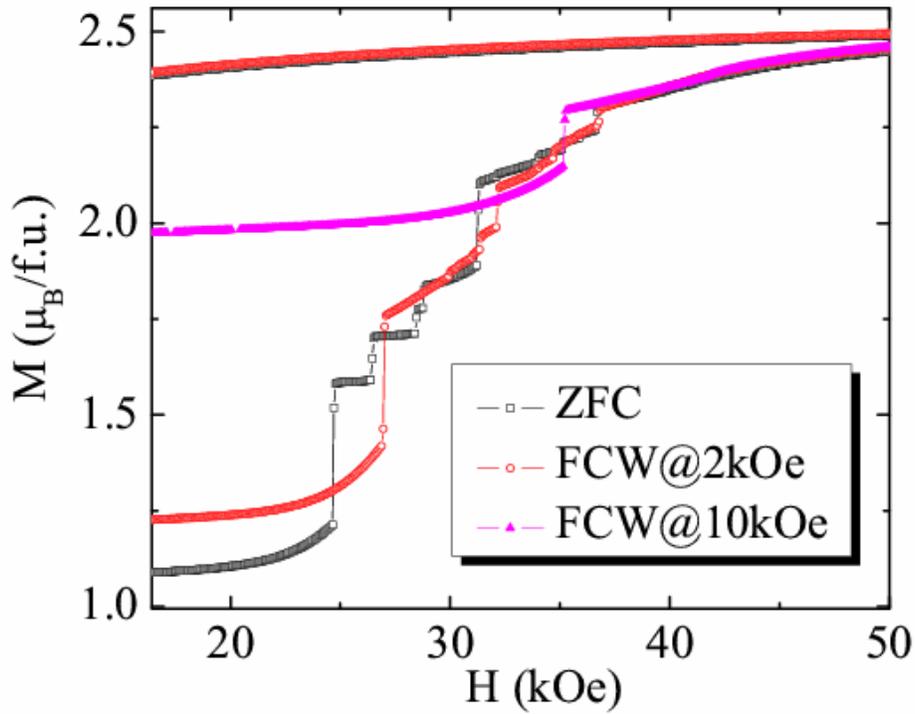

FIG. 6. Field dependence of ZFC and FCW magnetization at =1.8K in Ce(Fe$_{0.975}$Ga$_{0.025}$)$_2$. During FCW measurement, the cooling field was made to zero at the measurement temperature then data was taken for the increasing and decreasing field cycles.

The dependence of the magnetization on the cooling field was also studied in these compounds. Fig. 6 shows the field dependence of the ZFC and FCW magnetization at =1.8 K. The sample was heated above 240K (> $T_C$) and then cooled to 1.8 K for each measurement. In the FCW mode, the cooling field was reduced to zero at the measurement temperature and the magnetization was measured subsequently. Comparing the ZFC and FCW data, it may be noted that the region of existence of antiferromagnetic phase is the least in the ZFC case and increases with the cooling field in the FCW case. Furthermore, it can be seen that the number of steps in the *M-H* curve has decreased in the FCW curves, as compared to the ZFC curve. Application of fields higher than 10 kOe may cause the sample to be converted to fully ferromagnetic state. Another point worth noting from this figure is that the sharp steps are shifted towards higher fields when the sample is field cooled. This kind of behavior is also observed in perovskite manganites[19]. This implies that the AFM-FM co-existence region has broadened by increasing the cooling field. This is surprising because while magnetic field enhance the parallel coupling of the moments, here AFM exchange interaction is also getting enhanced although magnetization value is increased with increase in the cooling field.

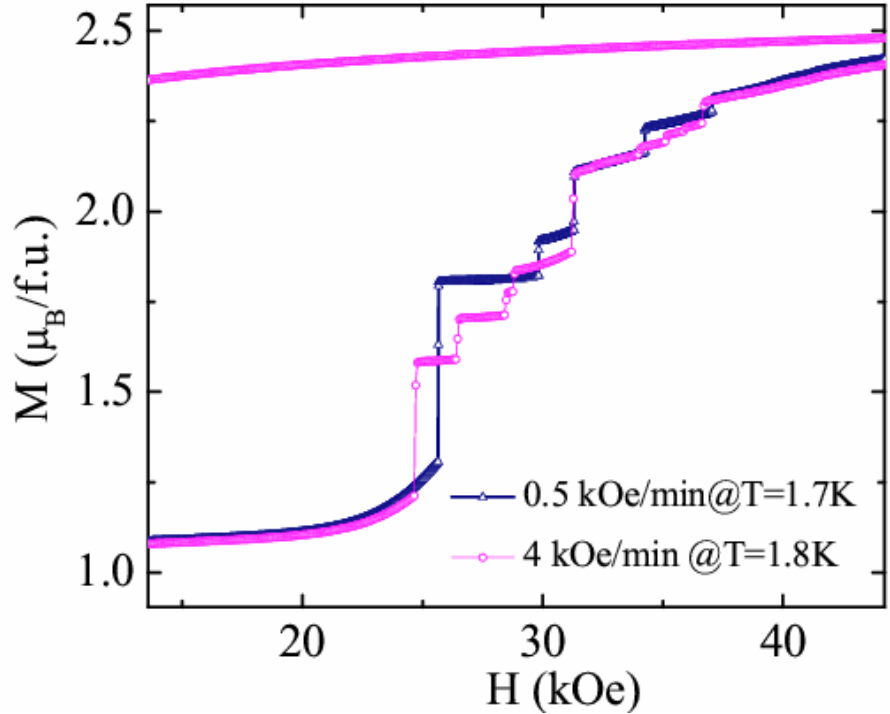

FIG. 7. Two loop *M-H* isotherms measured for Ce(Fe$_{0.975}$Ga$_{0.025}$)$_2$ at T=1.8 K with a field sweep rates of 0.5 kOe/min. and 4 kOe/min. The sample was zero field cooled from 240 K before each measurement.

The effect of changing the sweep rate of the field on the magnetization behavior has also been investigated. Fig. 6 shows the M-H plots of Ce(Fe$_{0.975}$Ga$_{0.025}$)$_2$ at T=1.8 K with a field sweep rates of 0.5 kOe/min. and 4 kOe/min. Here also, the sample was zero field cooled from 240K before each measurement. It is interesting to see that when the sweep rate is very slow the steps occur at higher fields, as compared to that at faster sweep rates. It may be recalled here that with increase in temperature, the steps in *M-H* curve shift towards higher field and also the number of steps decreases.

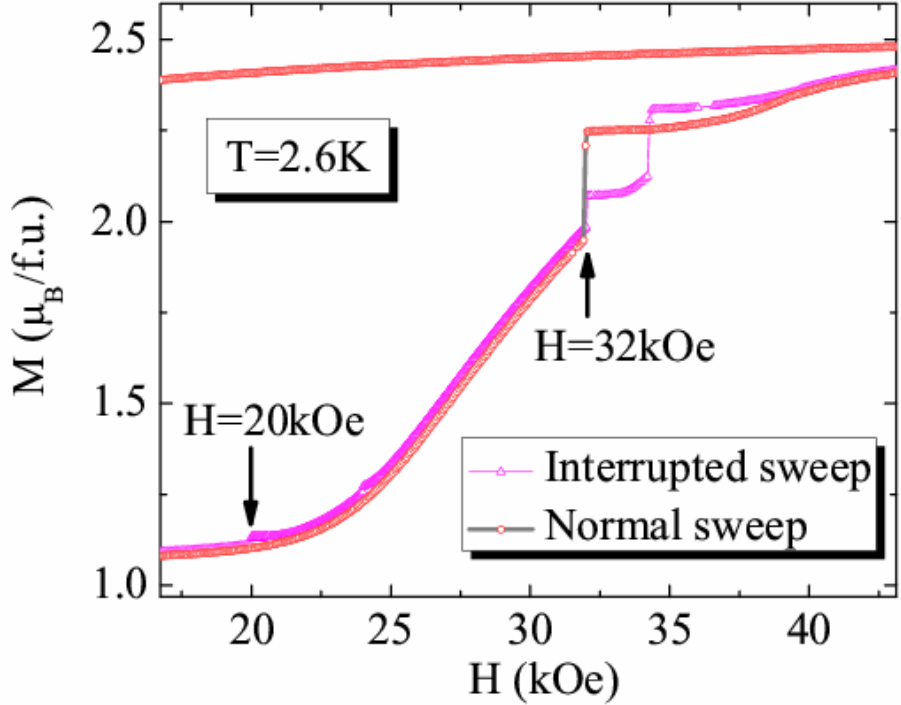

FIG. 8. Isothermal ZFC magnetization at T=2.6K of Ce(Fe$_{0.975}$Ga$_{0.025}$)$_2$ in a typical sweep and in an interrupted sweep. In interrupted sweep, fields of 20kOe and 32kOe were held constant for 1.5 hrs. and 1 hr.

Fig. 8 compares the *M-H* curves obtained in the normal sweep and interrupted sweep. In normal sweep, the data has been taken continuously in time as the field was increased from zero to 70kOe. A step at about 32 K was found in this curve. In the interrupted sweep, 20kOe was maintained for 1.5 hrs. and 32kOe was maintained for 1 hr. A step is observed at 20kOe when the field was held for 1.5 hrs., which was not observed in the normal sweep. At 32 kOe, when it was held for 1 hr., two steps with a shift in the field are found. During these holding times, the magnetization evolves in small steps to its final value (Fig. 9). The step sizes are even smaller at H=32 kOe, as shown in Fig. 9a and b. Insets in these figures show the magnetization steps during the holding time.

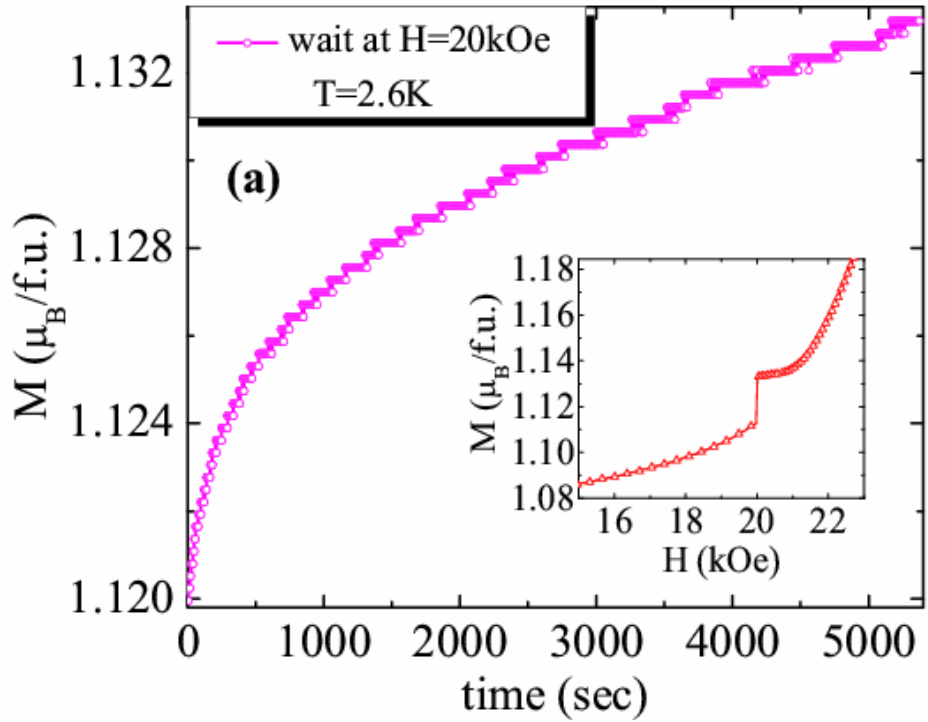
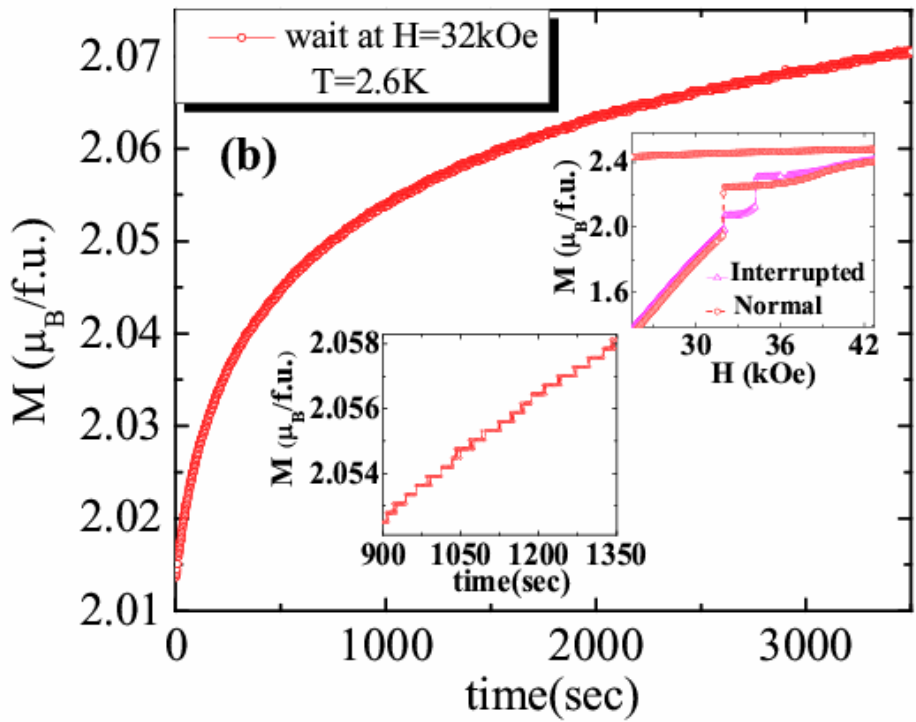

FIG. 9. Time evolution of isothermal (ZFC) magnetization at 2.6 K for Ce(Fe$_{0.975}$Ga$_{0.025}$)$_2$ during the holding time at (a) 20kOe and (b) 32kOe. Insets show the magnetization steps during the holding time.

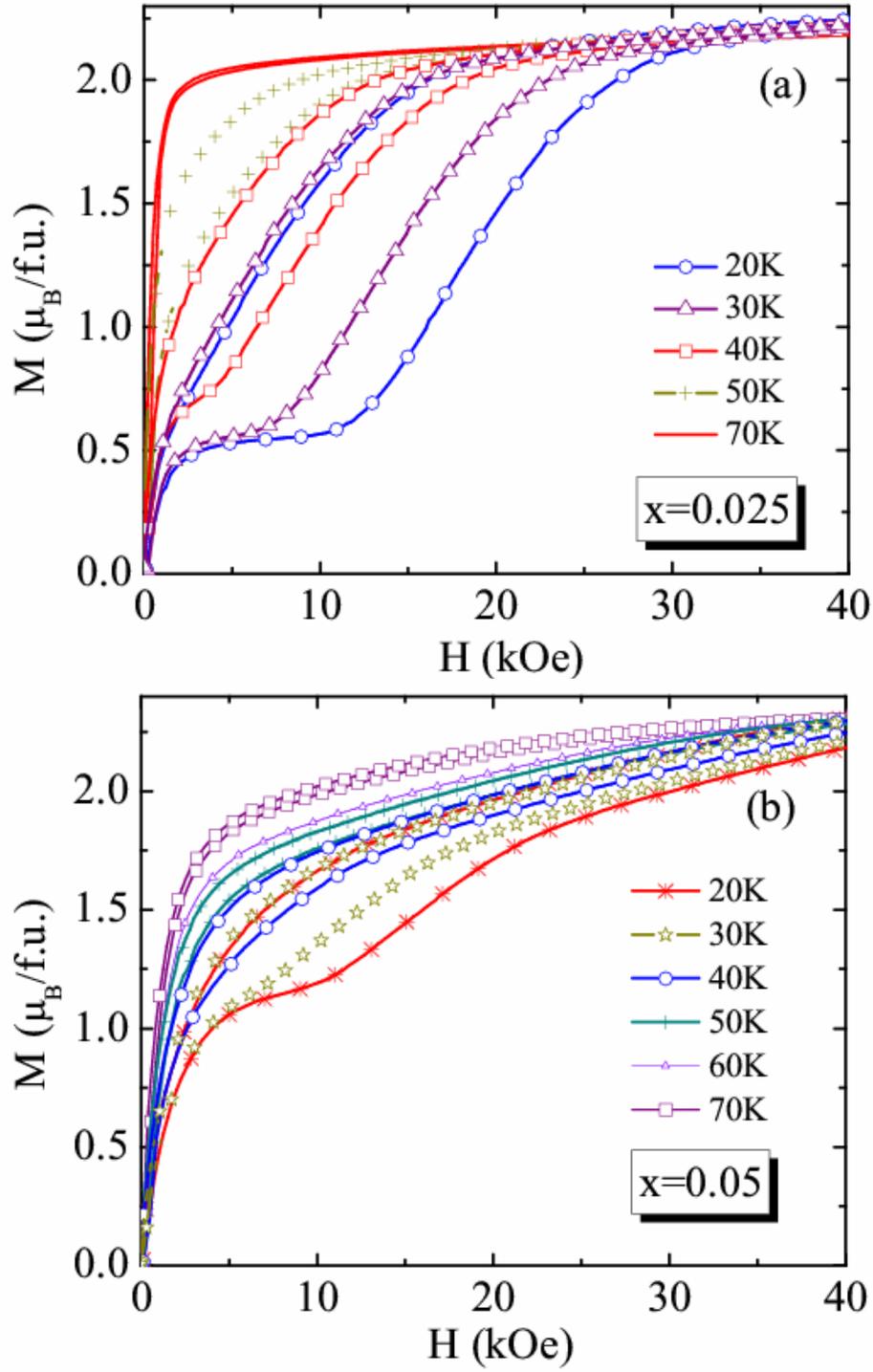

FIG. 10. Isothermal magnetization at high temperatures ($\geq$ 20K) in (a) Ce(Fe$_{0.975}$Ga$_{0.025}$)$_2$ and (b) Ce(Fe$_{0.95}$Ga$_{0.05}$)$_2$.

Fig. 10 demonstrates the nature of high temperature *M-H* isotherms of the compounds with *x*=0.025 and *x*=0.05. The transition is rather continuous and the critical field for the growth of FM phase is about 14kOe which is less compared to $H_{c1}$ ~ 24kOe obtained at 2.5K. Further increase in temperature makes the transition broader and the area of the hystersis loop also becomes smaller. The kink in the low field, indicating the growth of FM phase has disappeared above 45K for *x*=0.025 sample. For *x*=0.05, the transition is not as sharp as in *x*=0.025 sample, as we have seen earlier. The AFM-FM transition temperatures for these compounds are around 50K, as found from the *M-T* curve [Fig. 2]. So around and above this temperature the metamagnetic type transition gets diminished and we get a typical ferromagnetic type *M-H* plot.

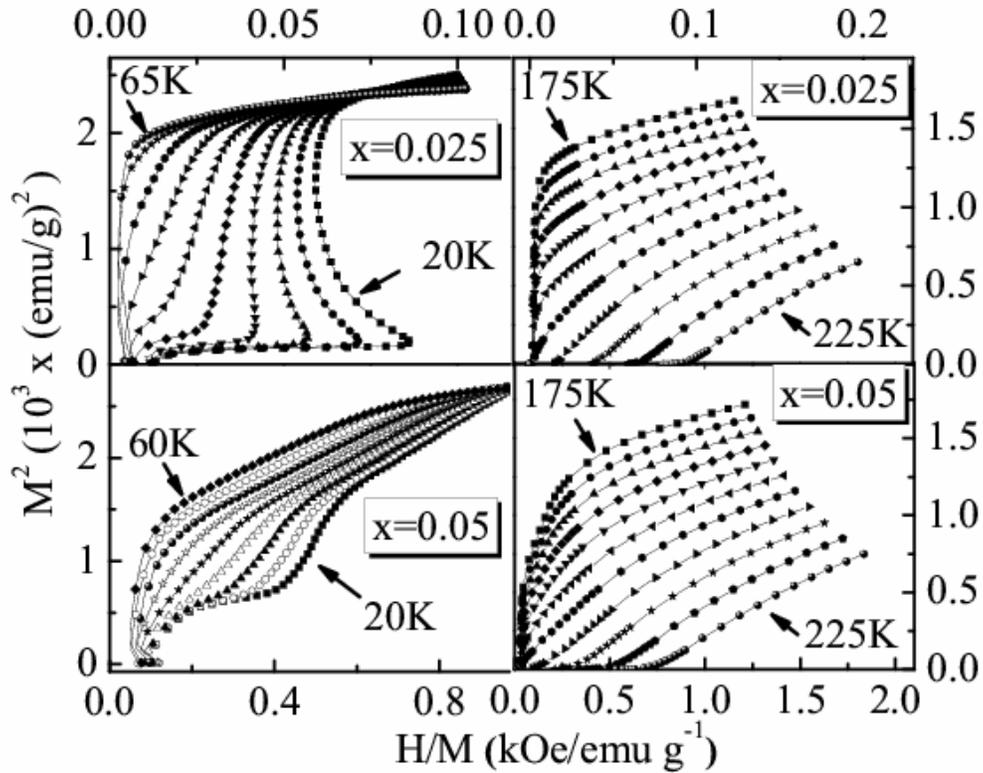

FIG. 11. Arrott's plots for x=0.025 and x=0.05 compounds at AFM-FM transition region and FM-PM transition region.

In order to understand the nature of the AFM-FM and FM-PM transitions, the Arrott plots have been recorded the compounds with $x$=0.025 and 0.05 and are shown in Fig. 11. It is quite clear that while the low temperature transitions are first order in nature, the high temperature ones are of second order. Also, the strength of first order is more in the case of $x$=0.025, as compared to $x$=0.05.

Based on the *M-H* curves obtained at various temperatures the magnetic phase diagram has been constructed for the compounds with $x$=0.025 and 0.05 and is shown in Fig. 12. In this figure, $H_{c1}$ and $H_{c2}$ refer to the lower and upper critical fields.

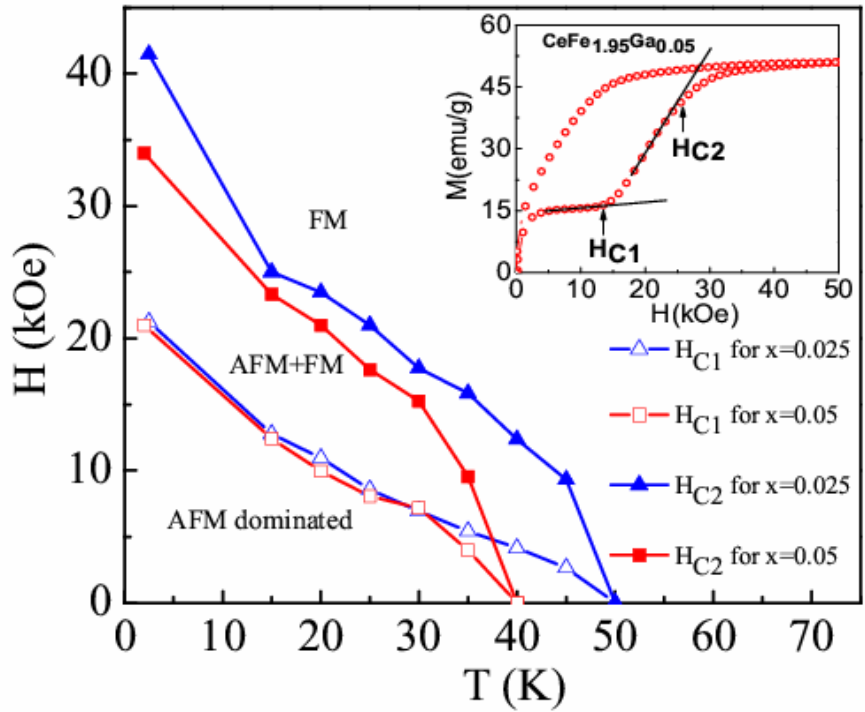

FIG. 12. Magnetic phase diagram for Ce(Fe$_{1-x}$Ga$_x$)$_2$ compounds with $x$=0.025 and 0.05

## IV. Discussions

The major findings from the results presented in the above section are (i) stabilization of low temperature antiferromagnetic phase with Ga substitution, (ii) sharp jumps in the magnetization curves in the Ga-substituted compounds across the AFM-FM transition, (iii) the fact that the virgin curve lies outside the envelope curve and (iv) first order AFM-

FM magnetic transition (v) the similarity between these compounds and the martensitic systems with regard to the magnetization relaxation behavior.

As mentioned earlier, recent studies on compounds showing colossal magnetoresistance (CMR) show ultrasharp steps when measurements are performed below 5K[8]. Such steps also observed in both single and polycrystalline samples[14]. The mechanism behind such steps has been attributed to the catastrophic relief of strain built up during the first order martensitic phase transition. Magnetocaloric material $Gd_5Ge_4$ also shows similar type of sharp step across AFM-FM transition[18]. It has been proposed that the temperature[20] and the magnetic-field-induced[8] phase evolution in these systems resembles that of a martensite. In the present case, the FM phase grows above 50K for $x=0.025$ and $x=0.05$ samples. It is well documented that the magnetic phase transition in these kinds of compounds is also associated with structural transition from rhombohedral AFM phase to cubic FM phase[17, 21-23]. The application of magnetic field at low temperature favors the FM phase, and as the field is increased, the volume ratio of FM phase increases and results in a catastrophic release of strain associated with the magneto-structural transition. This must be responsible for the magnetization steps. Till now, only Ru and Re doped $CeFe_2$ are reported to show the strain-induced first order magnet-structural transition and therefore, the present study gives yet another evidence for this anomalous behavior of the $CeFe_2$-based systems.

The sharpness of the transitions makes us to think of martensitic type nature in this system. As per literature, the AFM phase is rhombohedral and FM phase is cubic [17,21-23]. So this a phase separated system where FM phase can be induced by applying field externally. While doing so it will be associated with a huge strain relief during the transformation between AFM phase and FM phase. The sweep rate dependence also confirms this assumption. A slower sweep is supposed to transform these two phases smoothly compared to higher sweep rate. A progressive accommodation of martensitic strains during the field induced order-order transition leads to higher critical field which is needed for the transition to take place.

The relaxation effect seen in these magnetization data points towards a glassy phase at low temperature, for these compounds. Comparing the *M vs. T* behavior at different fields for Ce(Fe$_{0.975}$Ga$_{0.025}$)$_2$ sample, it is observed that at higher field like 10kOe the difference between ZFC and FCW data is very large compared to the data taken at field of 0.5kOe. This implies strong magnetic frustration in this system, at low temperatures. It is also observed that the field cooling shifts the magnetization steps to higher fields. It is to be noted that the cooling field is well below the critical fields corresponding to the magnetization jumps. This may be due to the creation of a new magnetic coupling scheme with different interaction energies between two neighboring moments, in the FCW mode. That interaction energy is probably larger compared to the ZFC scenario, which requires higher fields to cause the metamagnetic transitions. Much larger cooling field may be able to suppress the AFM phase completely and render the system a simple ferromagnet.

The relaxation results show that the present system is similar to some manganite systems, which relax its moments when left idle for some time in certain magnetic field. The steps can evolve in time, even when keeping the field and the temperature constant. The geometry of the lattice can cause magnetic frustration. In three dimensions the well-known frustrated system is the pyrochlore structure, in which the magnetic ions occupy a lattice of corner sharing tetrahedra. The low temperature AFM structure of doped CeFe$_2$ is similar to this pyrochlore structure. This type of structure leads to a situation in which there can be no single unique ground state, but a variety of similar low energy states, resulting in frustration.

## V. Conclusions

The present study shows that Ga doping stabilizes the dynamic AFM state in CeFe$_2$. The FM-AFM transition is found to be of first order in nature. We find that phenomena such as strain induced first order jumps in the magnetization curves, asymmetry between the *M-H* curves during the increasing and decreasing field cycles, the fact that the envelop curve is inside the virgin curve occur in these compounds as well, like the Ru and Re-doped $CeFe_2$. The existence region of the AFM phase is found to increase considerably with field cooling. Thermomagnetic history is found to influence the magnetization behavior. It is found that the magnetization steps can be induced by proper relaxation procedure. Experimental evidences clearly show that the system is frustrated at low temperatures. Finally, the results, in general, show that Ga- doped $CeFe_2$ shows a martensitic-like behavior due to the strong magneto-structural coupling.